\documentclass[onecollarge,natbib]{svjour2}
\bibpunct{[}{]}{;}{n}{}{,} 
\smartqed  
\usepackage{graphicx}
\newcommand{\nn}{\nonumber}        
\newcommand{\bm}[1]{\mbox{\boldmath $#1$}}

\newcommand{\open}{{<\kern -0.3 em{\scriptscriptstyle )}}}

\newcommand{\nslash}{\kern 0.2 em n\kern -0.45em /}
\newcommand{\Pslash}{\kern 0.2 em P\kern -0.56em \raisebox{0.3ex}{/}}
\newcommand{\pslash}{\kern 0.2 em p\kern -0.4em /}
\newcommand{\kslash}{\kern 0.2 em k\kern -0.45em /}
\newcommand{\Sslash}{\kern 0.2 em S\kern -0.56em \raisebox{0.3ex}{/}}

\newcommand{\eq}{\begin{equation}}
\newcommand{\ee}{\end{equation}}
\newcommand{\beq}{\begin{equation}}
\newcommand{\eeq}{\end{equation}}
\newcommand{\ba}{\begin{eqnarray}}
\newcommand{\ea}{\end{eqnarray}}

\newcommand{\sumint}{\kern 0.2 em {\textstyle\sum} \kern -1.1 em \int}

\newcommand{\Tr}{{\rm Tr}}
\journalname{Few-Body Systems}
\begin{document}

\title{Average transverse momentum quantities approaching the lightfront}

\author{Dani\"el Boer}


\institute{D. Boer \at 
Van Swinderen Institute, University of Groningen\\
Nijenborgh 4, NL-9747 AG Groningen, The Netherlands\\
                \email{d.boer@rug.nl} }

\date{September 29, 2014} 

\maketitle

\begin{abstract}
In this contribution to Light Cone 2014, three average transverse momentum quantities are discussed: the Sivers shift, the dijet imbalance, and the $p_T$ broadening. The definitions of these quantities involve integrals over all transverse momenta that are overly sensitive to the region of large transverse momenta, which conveys little information about the transverse momentum distributions of quarks and gluons inside hadrons.  
TMD factorization naturally suggests alternative definitions of such integrated quantities, using Bessel-weighting and rapidity cut-offs, with the conventional definitions as limiting cases. The regularized quantities are given in terms of integrals over the TMDs of interest that are well-defined and moreover have the advantage of being amenable to lattice evaluations.

\end{abstract}

\section{Introduction}
\label{intro}
Many measurements of processes sensitive to the transverse momentum of partons have been performed. 
For example, semi-inclusive DIS (SIDIS), $e\, p \to e' h X$,  has been studied as a function of 
the transverse momentum $\bm{P}_{h\perp}$ of the final state hadron $h$ with respect to the 
$\gamma^* p$ axis. In a tree level analysis one can relate this observed transverse momentum to 
the ``intrinsic'' partonic transverse momentum, but higher order corrections will modify this relation. 
The transverse momentum distribution can have a correlation with the spin 
$S_T$ of the initial proton and with the spin of the quark, the so-called Sivers \cite{Sivers:1989cc} and 
Boer-Mulders \cite{Boer:1997nt} effects. They lead to transverse momentum dependent azimuthal 
modulations of cross sections. Such azimuthal asymmetries are most naturally described in terms 
of transverse momentum dependent parton distribution functions (TMDs). Average 
transverse momentum quantities, such as $\langle k_T \times S_T\rangle$ (related to the Sivers shift) 
and $\langle k_T^2\rangle$, can be used to quantify the width of the various transverse momentum distributions.  
Considering such average quantities within the TMD framework has its complications, however, 
such as possible divergences and process dependence. It has become clear since 2002 that TMDs in general 
need not be universal, due to gluon rescattering effects \cite{Brodsky:2002cx,Collins:2002kn,Belitsky:2002sm}. 
Summation of all gluon insertions leads to path-ordered exponentials in the TMD correlators \cite{Efremov:1978xm}.
The resulting Wilson lines depend on whether the color is incoming or outgoing \cite{Collins:1983pk,Boer:1999si,
Collins:2002kn,Brodsky:2002rv,Belitsky:2002sm,Boer:2003cm}. Although this does not automatically imply that this 
will affect observables, it turns out that it does in certain cases, for example, in Sivers effect asymmetries.
The gauge invariant definition of the Sivers TMD ($f_{1T}^\perp$) is given by a Fourier transform (F.T.) of a hadronic 
matrix element of a nonlocal operator,  
 \beq 
 P \! \cdot \! \left({\bm k}_T \times {\bm S}_T \right) {f_{1T}^{\perp [{\cal C}]}(x,{{\bm k}_T^2})} \propto\ {\rm F.T.} \, \langle P, {S_T}| \bar{\psi}(0) \, {\cal L}_{{\cal C}[0,\xi]} \, \gamma^+ \, \psi(\xi) |P, {S_T}\rangle\big|_{\xi = (\xi^-,0^+,{\bf \xi}_T)}, 
\eeq 
which contains a gauge link, which is a Wilson line along a path ${\cal C}$ from $0$ to $\xi$:
\beq
{\cal L}_{{\cal C}}[0,\xi] = {\cal P} \exp \left(-ig\int_{{{\cal C}[0,\xi]}} 
ds_\mu \, A^\mu (s)\right).
\eeq
In semi-inclusive DIS it has a future pointing staple-like Wilson line 
arising from final state interactions (FSI), whereas in Drell-Yan (DY) it is past pointing from initial state interactions (ISI).
One can use parity and time reversal invariance to relate the Sivers functions of SIDIS and DY, yielding the overall sign relation:  
$f_{1T}^{\perp [{\rm SIDIS}]} = - f_{1T}^{\perp [{\rm DY}]}$ \cite{Collins:2002kn}. Experimental studies to test this relation are ongoing at CERN (COMPASS) \& Fermilab (SeaQuest), and are considered at several other places, e.g.\ NICA at JINR and J-PARC.This is not just a test of this relation, but of the TMD formalism as a whole. 

Gauge links track the color flow in a process. The more hadrons are observed in a process, the more complicated 
the end result, yielding more complicated $N_c$-dependent pre-factors \cite{Bomhof:2004aw,Buffing:2012sz,Buffing:2013dxa}. 
Too many color flow directions can even lead to factorization breaking \cite{Collins:2007nk,Rogers:2010dm}.

All this is not just a peculiarity of polarized protons, it also affects scattering of unpolarized hadrons. For example, it affects the muon pair angular distribution in the DY process through the double Boer-Mulders effect 
\cite{Boer:1999mm,Boer:2002ju,Buffing:2013dxa} and also the transverse momentum distribution in Higgs production at the LHC \cite{Sun:2011iw,Boer:2011kf} through the linear polarization of gluons inside unpolarized protons \cite{Mulders:2000sh}.

QCD corrections will also attach to the Wilson line, which as a consequence needs renormalization. 
The Wilson lines are not smooth and involve the so-called cusp anomalous dimension \cite{Polyakov:1980ca,Dotsenko:1979wb,Korchemsky:1987wg}. 
As a regularization, in TMD factorization the path can be taken off the lightfront, specified by 
a rapidity cut-off parameter $\zeta$. In short, the TMDs are not just functions $f(x,k_T^2)$ of the momentum fraction $x$ and the transverse momentum $k_T$, 
but rather $f^{[{\cal U}]}(x,k_T^2;\mu,\zeta)$, where $\mu$ denotes the renormalization scale and ${\cal U}$ the Wilson line. 
These complications turn out to be also advantageous. First of all, the change with $\mu$ and $\zeta$ yields the energy evolution of 
TMD observables (for the Sivers asymmetries this appears to work \cite{Aybat:2011ta,Sun:2013dya,Echevarria:2014xaa}, but needs to be tested over a larger energy range). 
It shows that the TMDs and the corresponding asymmetries generally become broader and smaller with increasing energy. At high scales 
they develop power law tails \cite{Aybat:2011zv}, which matter much in integrals over all transverse momenta, as will be discussed. 
Another advantage is that finite $\zeta$ allows for the calculation of the Sivers effect on the lattice (quite unexpectedly).

\section{Transverse momentum weighting}

The extraction of TMDs is complicated because TMDs appear in convolution expressions that appear in different ways in different processes. 
Integration weighted with powers of the observed transverse momentum $P_{h\perp}$ was suggested as a solution, as it projects 
out ``portable'' functions \cite{Kotzinian:1997wt,Boer:1997nt}.
In this way weighted asymmetries become expressions in terms of transverse moments of TMDs:
\beq
f^{(n)}(x) \equiv \int d^2 \bm{k}_T \left({\frac{\bm{k}_T^2}{2M^2}}\right)^n f(x,\bm{k}_T^2)
\eeq
Such transverse moments appear in different processes in exactly the same form
(except possibly for a calculable $N_c$-dependent prefactor). The $n=1$ moment of the Sivers function 
appears in the Sivers shift, which is the average transverse momentum shift orthogonal to a given transverse polarization~\cite{Boer:2011xd}:
\beq
\langle k_{T}^y(x) \rangle_{TU} = \left. \frac{ \int d^2 \bm{k}_T\, \bm{k}_T^y  \ \Tr\left[ \Phi(x,\bm{k}_T;P,S;\mu,\zeta) \gamma^+\right]}{ \int d^2 \bm{k}_T\, \Tr\left[ \Phi(x,\bm{k}_T;P,S;\mu,\zeta) \gamma^+\right]} \right|_{S^\pm=0=S_T^y,\, S_T^x=1} = M \frac{ f_{1T}^{\perp(1)}(x;\mu,\zeta) }{ f_1^{(0)}(x;\mu,\zeta)} .
\eeq
First a comment on the function $f_1^{(0)}(x;\mu,\zeta)$ in the denominator. Because of the rapidity parameter $\zeta$, the integral over all transverse momenta need not yield the collinear function $f_1(x;\mu)$ exactly \cite{Collins:2003fm}, $\int d\bm{k}_T f_1(x,\bm{k}_T; \mu,\zeta) \stackrel{?}{=}  f_1(x;\mu)$, similar to the old question whether $f_1(x;Q) \stackrel{?}{=} \int_{|k_T|< Q} d^2 k_T\, f_1(x,k_T^2)$. 
Instead of such relations where the TMD determines the collinear function, it is rather the 
collinear function that determines the large transverse momentum part of the TMD, the perturbative tail:
\beq
f_1(x,\bm{k}_T^2) \stackrel{\bm{k}_T^2 \gg M^2}{\sim}   \alpha_s\,\frac{1}{\bm{k}_T^2} \, \left(K \otimes f_1\right) (x).
\eeq
The transverse momentum integration of this expression diverges if not suitably cut off. Similarly, without bothering about scale dependences, the $n=1$ moment of the Sivers function is related \cite{Boer:2003cm} to the collinear twist-3 Qiu-Sterman function $T(x,S_T)$ \cite{Qiu:1991pp}:
\beq
f_{1T}^{\perp {(1)}[\pm]}(x) \equiv \int d^2 \bm{k}_T {\frac{\bm{k}_T^2}{2M^2}} f_{1T}^{\perp [\pm]} (x, \bm{k}_T^2) = \mp \frac{T(x,S_T)}{2M} ,
\eeq
where in $A^+=0$ gauge
\beq
{T(x,S_T)} = i \frac{g M}{P^+} \int \frac{d \lambda}{2\pi} e^{i\lambda x}\langle P,S|\,  \overline \psi (0) {\Gamma_\alpha} 
{\int d\eta \;  F^{+\alpha} (\eta n_-)}\; \psi(\lambda n_-)\, | P,S\rangle
\eeq
with ${\Gamma_\alpha} \equiv  \epsilon_{T \beta \alpha} S_T^{\beta} \not\! \!n_-/(2i M P^+) $. Again due to the $\zeta$ dependence this relation between Sivers and Qiu-Sterman functions is not well-determined. Again conversely it is the perturbative tail of the Sivers function that is determined by the Qiu-Sterman function \cite{Ji:2006ub,Koike:2007dg}
\beq
f_{1T}^\perp(x,\bm{k}_T^2) \stackrel{\bm{k}_T^2 \gg M^2}{\sim} \alpha_s\, \frac{M^2}{\bm{k}_T^4} \, \left(K' \otimes T_F\right) (x).
\eeq
Integrating over this perturbative tail without cut-off or regulator will yield a divergent result for the first tranverse moment of the Sivers function. 

Conventional weighting with powers of transverse momentum assumes that 1) the integrals converge or are suitably regulated somehow, and 2) taking
integrals over only TMD expressions (which are only valid for $Q_T = |\bm{P}_{h\perp}|/z_h \ll Q$) is fine. To by-pass these tricky issues, both connected with the perturbative tails, one can consider instead a modified weighting: Bessel weighting \cite{Boer:2011xd}, where instead of $|P_{h\perp}|^n$ one uses $J_n(|P_{h\perp}|{\cal B}_T) n! \left(2/{\cal B}_T\right)^n$. If ${\cal B}_T$ is not too small, the contribution from the perturbative tail at high transverse momentum will be suppressed and the TMD region ($Q_T \ll Q$) should dominate. In the limit ${\cal B}_T \to 0$, when conventional weighting is retrieved, the perturbative tails become very important and divergences may arise. 
Application of the Bessel-weighting method to asymmetries is discussed in detail in \cite{Aghasyan:2014zma}.
Bessel-weighted asymmetries involve generalized transverse moments, called Bessel moments for short:
\beq
\tilde f^{(n)}(x, \bm{b}_T^2 ) = \frac{ 2\pi \ n!}{(M^2)^n} \int  d |\bm{k}_T| |\bm{k}_T|\left( \frac{|\bm{k}_T|}{|\bm{b}_T|}\right)^n J_n(|\bm{b}_T||\bm{k}_T|)\  f(x, {\bm{k}_T^2} ).
\eeq
The idea is that the convolution in terms of TMDs is best de-convoluted by Fourier transform, in which derivatives of TMDs appear naturally, e.g.\ for unpolarized hadrons: 
\begin{eqnarray}
\Phi(x,\bm{k}_T) & = & \frac{M}{2}\,\Biggl\{ {f_1(x ,\bm{k}_T^2)}\, \frac{\not \!\! P}{M} + {h_1^\perp (x,\bm{k}_T^2)} \, \frac{i \not\! k_T \not \!\! P}{M^2} \Biggl\}\nn\\
& \stackrel{{\rm F.T.}}{\longrightarrow} & \quad {\tilde{\Phi}}(x,\bm{b}_T)\ =\ \frac{M}{2}\,\Biggl\{ {\tilde{f}_1(x ,\bm{b}_T^2)}\, \frac{\not \! \! P}{M} + {\left(\frac{\partial}{\partial \bm{b}_T^2} \tilde{h}_1^\perp (x,\bm{b}_T^2) \right)} \, \frac{2 \not\! {\bm{b}_T} \not \! \!P}{M^2} \Biggl\}.
\end{eqnarray}
The conventional transverse moments can be viewed as limits of derivatives of TMDs: 
\beq
\tilde f^{(n)}(x, \bm{b}_T^2) = n!\left( -\frac{2}{M^2}\partial_{ \bm{b}_T^2} \right)^n \ \tilde f(x, \bm{b}_T^2) \stackrel{\bm{b}_T^2 \to 0}{\longrightarrow} f^{(n)}(x)
\eeq
In other words, such derivatives of TMDs are the generalized transverse moments, the Bessel moments, appearing in  
Bessel-weighted asymmetries. 

An additional advantage is that nonzero ${b}_T \equiv |\bm{b}_T|$ (and finite $\zeta$) allows for calculation of TMDs on the lattice. The Bessel-weighted analogue of the Sivers shift is \cite{Boer:2011xd}:
\begin{eqnarray}
\langle k_T^y(x) \rangle_{TU}^{{b}_T} &  = & \left. \frac{ \int d |\bm{k}_T|\, |\bm{k}_T| \int d\phi_p \frac{2J_1(|\bm{k}_T||\bm{b}_T|)}{|\bm{b}_T|}\, \sin(\phi_p-\phi_S)\, \Tr\left[ \Phi(x,\bm{k}_T;P,S;\mu,\zeta) \gamma^+\right]}{ \int d |\bm{k}_T|\,|\bm{k}_T| \int d\phi_p J_0(|\bm{k}_T||\bm{b}_T|)) \, \Tr\left[ \Phi(x,\bm{k}_T;P,S;\mu,\zeta) \gamma^+\right]} \right|_{S^\pm=0=S_T^y,\, S_T^x=1} \nonumber \\ 
&  =  & M \frac{ \tilde{f}_{1T}^{\perp(1)}(x,{b}_T^2;\mu,\zeta) }{ \tilde{f}_1^{(0)}(x,{b}_T^2;\mu,\zeta)}.
 \end{eqnarray}
For nonzero ${b}_T$ this involves well-defined (finite) quantities, with Wilson lines 
that are off the lightcone (spacelike) for finite $\zeta$. After taking Mellin moment, one has a well-defined quantity of $n$ and ${b}_T$ that can be and has been evaluated on the lattice \cite{Musch:2011er}. Although smaller pion masses and larger $\zeta$ should be considered, this lattice study constitutes the first `first-principles' demonstration within QCD that the Sivers and Boer-Mulders functions are nonzero. Moreover, it corroborates the overall sign relation between SIDIS and DY gauge links (as it should) and is consistent with the up Sivers function in SIDIS being negative and the down Sivers function positive, as obtained in models and fits to data.  

As $\zeta$ is taken larger the Wilson lines are approaching the lightfront. In case of the nucleon one should consider 
\beq
\hat{\zeta}=\frac{\zeta}{2 m_N} = \frac{\bm{v}\cdot \bm{P}}{\sqrt{|\bm{v}^{\, 2}|} \sqrt{P^2}} = \sinh(y_P-y_v) \gg \frac{\Lambda_{\rm QCD}}{2 m_N} \approx 0.1 
\eeq
To avoid large logs, optimal Collins-Soper parameter $\zeta = {\cal O}(Q) \gg m_N$. Recent studies of the pion Boer-Mulders shift have a much improved determination of the $\zeta$ dependence, which shows that for larger values it approaches a constant rather quickly \cite{Engelhardt:2014wra}. This is very promising for a stable determination of the large $\zeta$ Boer-Mulders and Sivers shifts. The other limit of interest, i.e.\ $b_T \to 0$, also appears to approach a constant fast for larger $\hat\zeta$ values.  
The limit ${b}_T \to 0$ of the Sivers shift can tell us something about the Qiu-Sterman function:
\beq
\lim_{b_T \to 0} \tilde{f}_{1T}^{(1)[+]}(x,b_T^2;\mu,\zeta)\stackrel{?}{=}-\frac{T(x,S_T;\mu)}{2M} ,  
\eeq
where the `?' is included because of the $\zeta$ dependence of the r.h.s., just like for:
\beq
\lim_{b_T \to 0} \tilde{f}_1^{(0)}(x,b_T^2;\mu,\zeta) \stackrel{?}{=} f_1(x;\mu). 
\eeq
Strictly speaking, such an identification is only meaningful when viewed as part of the full cross section expression in which the $\zeta$ dependence cancels, but if the limits of  
${b}_T \to 0$ and large $\zeta$ become constant, the limit of the Bessel moment may be unambiguous. It is a very interesting limit to consider, since the Qiu-Sterman function itself is intrinsically non-local along the lightcone (involving $\int \! d\eta^- \; F^{+\alpha}(\eta^-)$) and cannot be evaluated on the lattice.

\section{Average transverse momentum}

The dijet imbalance in hadron-hadron collisions is used to extract the average 
partonic transverse momentum. Dijets, and also dimuons and diphotons, have the advantage that there is no contribution from the 
fragmentation process. The observed transverse momentum $Q_T$ of a pair of jets will generally be different from that 
of a pair of muons or photons \cite{Begel:1999rc} because of the different partonic scattering contributions: 
$\langle Q_T^2 \rangle_{{\rm pair}}/2 = \langle k_T^2 \rangle_{{\rm intrinsic}}+ \langle k_T^2 \rangle_{{\rm soft}} + \langle k_T^2 \rangle_{{\rm NLO}}$. 
Dimuons/diphotons are mainly sensitive to 
$\langle k_T^2\rangle$ of the quarks, whereas dijets also probe the intrinsic transverse momentum of gluons. 
In addition, they are affected differently by spin correlation effects that are in principle also present \cite{Boer:2009nc,Pisano:2013cya}. 
For the idealized case of equal jet transverse momenta (both equal to $E_T/2$) the differential cross section of dijet production takes the form \cite{Boer:2009nc} 
\beq
\frac{d\sigma}{dE_T d\delta\phi}= A(Q_t^2) + B(Q_t^2) Q_t^2 +
C(Q_t^2) Q_t^4, 
\eeq
where $Q_t=E_T |\sin(\delta \phi/2)|$. Here $B$ comes from the double Boer-Mulders effect for quarks \cite{Boer:1999mm,Boer:2002ju,Lu:2008qu} and $C$ from the gluon analogue. The latter contribution is absent for dimuons or diphotons. Due to the presence of both  ISI and FSI, the dijet expressions may not factorize though~\cite{Rogers:2010dm}, which complicates or may even prevent the extraction of $\langle k_T^2\rangle$ from the dijet imbalance in hadron-hadron collisions. 

The average $Q_T^2$ can be measured, but how is the $\langle k_T^2 \rangle$ defined here? Higher transverse moments of $f_1$, starting with $\langle k_T^2 \rangle = 2 M^2 f_1^{(1)}$, generally diverge 
due to the power law tail. The average $k_T$ can be defined by a Gaussian fit to the low $k_T$ part of the distribution as done in \cite{Aybat:2011zv}, but as the energy increases this becomes less accurate and less relevant. We propose a definition using Bessel weighting instead:
\beq
\frac{\langle k_T^2 \rangle}{2 M^2} = f_1^{(1)}(x)
\to \tilde f_1^{(1)}(x, \bm{b}_T^2) = \frac{2\pi}{M^2} \int  d |\bm{k}_T| \frac{|\bm{k}_T|^2}{|\bm{b}_T|} J_1(|\bm{b}_T||\bm{k}_T|)\  f_1(x, {\bm{k}_T^2} ).
\eeq
With respect to cutting off the perturbative tail any regularization will do, but 
Bessel weighting is natural from the perspective of deconvoluting TMD expressions.
It suggests and allows a lattice study of the gauge link dependence of $\tilde{f}_1^{(1) {[{\cal U }]} }(x, \bm{b}_T^2) $. It can be shown that  
${\cal U}=+$ (SIDIS) and ${\cal U}=-$ (DY) are the same now, requiring consideration of TMD-factorizing processes with more complicated links~\cite{Buffing:2012sz}.

The difference of the average $p_T^2$ in scattering off a nucleus $A$ and a proton $p$ is called the ``$p_T$ broadening'' $\Delta p_T^2 \equiv \langle p_T^2 \rangle_A - \langle p_T^2 \rangle_p$, where $p_T$ denotes the observed transverse momentum. It quantifies the broadening of the $p_T$ spectrum that one expects from multiple scattering that becomes more frequent with increasing $A$. Relating the observable broadening to the analogous quantity at the quark level, $\Delta k_T^2$, leads to a difference of two in principle divergent quantities, due to the perturbative tail $\propto 1/k_T^2$. Although that difference will be finite, it converges very slowly to the ÔtrueÕ value as function of a cut-off on $k_T$.  
Here it is suggested to consider Bessel-weighted TMDs which are finite by themselves, such that their difference approaches $\Delta k_T^2$ in the small $b_T$ limit: 
\beq
\tilde{f}_1^{(1)q/A [{\cal U }]}(x, \bm{b}_T^2) - \tilde{f}_1^{(1)q/p[{\cal U }]}(x, \bm{b}_T^2) \stackrel{\bm{b}_T^2\to 0}{\longrightarrow} \Delta k_T^2{}^{[{\cal U }]}  \equiv \langle k_T^2 \rangle_A^{[{\cal U }]} - \langle k_T^2 \rangle_p^{[{\cal U }]}.
\label{kTbroadening}
\eeq
The limit of $b_T \to 0$ converges very slowly for a perturbative tail $1/k_T^2$, like for any other cut-off on this tail.   
In contrast, the observed $p_T$ distribution falls off much faster than $1/p_T^2$ and will not have this problem.
Using the Bessel moments, a well-defined ratio can also be formed, but as $b_T$ gets smaller the interesting information about the $A$ versus $p$ difference will be quickly lost, reaching $(\infty+\Delta)/\infty =1$:
\beq
R_{\Delta} \equiv \frac{\tilde{f}_1^{(1)q/A}(x, \bm{b}_T^2)}{\tilde{f}_1^{(1)q/p}(x, \bm{b}_T^2)} \stackrel{\bm{b}_T^2\to0}{\longrightarrow} 1.
\eeq
The quark broadening $\Delta k_T^2$ in principle depends on the gauge link ${\cal U}$, possibly leading to non-universal nuclear effects. Experimental study of such effects will be hard, as the $p_T$ broadening in different processes is different anyway, like in the collinear twist-4 treatment \cite{Xing:2012ii}, 
because of the different partonic subprocesses. It will be hard to disentangle that from the additional process dependence from ISI and/or FSI. The Bessel-weighting version of $\Delta k_T^2$ would allow a lattice study of the possible link dependence of the quark broadening, if one can create reliable nuclear states on the lattice (cf.\ e.g.~\cite{Detmold:2012eu}).

\section{Conclusions}
Bessel-weighted expressions are well-defined and emphasize the TMD region, as opposed to conventional average transverse momentum quantities. Bessel-weighted TMDs, including the process-dependent Sivers function, are calculable on the lattice for finite $b_T$ and $\zeta$. 
The limit $b_T \to 0$ should be taken with care, because divergences and operator mixing typically arise, but in case it is a stable limit, the lattice result can even tell us about the size and shape of the Qiu-Sterman function. 

Extraction of the average partonic transverse momentum from experimental observables is generally very complicated due to spin correlation effects and multiple sources of process dependence. As a result, the commonly used observable like the dijet imbalance may actually not allow for extraction of the average partonic transverse momentum, as opposed to for instance the dimuon and diphoton imbalance.
The average transverse momentum and its broadening with atomic number $A$ for nuclei can also be redefined in a well-defined manner by considering Bessel weighting, allowing them to be studied on the lattice. 
The possible relation between the process and $A$ dependence of $f_1$ can be studied in this way, as the average transverse momentum and its $A$ dependence need not be universal.

\end{document}